\documentclass[letterpaper, 10 pt, conference]{ieeetran}  

\IEEEoverridecommandlockouts                              

\usepackage{amsmath} 

\newtheorem{Lemma}{Lemma}
\newtheorem{Theorem}{Theorem}

\title{\LARGE \bf
The inverse cyclotomic Discrete Fourier Transform algorithm
}

\author{Sergei V. Fedorenko$^{1}$
\thanks{*The reported study was funded by RFBR according to the research project No. 16-01-00716 a.}
\thanks{$^{1}$Sergei V. Fedorenko is with National Research University Higher School of Economics, Russian Federation 
        {\tt\small sfedorenko@hse.ru}}%
}

\begin{document}

\maketitle
\thispagestyle{empty}
\pagestyle{empty}

\begin{abstract}

The proof of the theorem concerning to the inverse cyclotomic Discrete Fourier Transform algorithm 
over finite field is provided.

\end{abstract}

\section{INTRODUCTION}

The discrete Fourier transform (DFT) can be applied in error correcting codes and code-based cryptography. 
The cyclotomic DFT method \cite{Fedorenko03,Fedorenko04} is the best one for computing the DFT over finite field. 
Alexey Maevskiy pointed that formula \hbox{\cite[(6)]{Fedorenko04}} in the example of paper \cite{Fedorenko04} 
had not been proved. 
We have corrected this mistake and introduced the proof.

\section{BASIC NOTIONS AND DEFINITIONS}

The DFT of length ${n \mid 2^m-1}$ of a vector 
$f = (f_i)$, \, $i \in [0,n-1]$, $f_i \in GF(2^m)$,
is the vector $F = (F_j)$
$$F_j = \sum_{i=0}^{n-1} f_i \alpha^{ij}, \quad j \in [0,n-1],$$ 
where $\alpha$ is an element of order $n$ in $GF(2^m)$. 
Let us write the DFT in matrix form

\begin{equation}
F = Wf,
\end{equation}
where $W = (\alpha^{ij})$, \, $i,j \in [0,n-1]$, is a Vandermonde matrix.
We assume that the length of the $n$-point Fourier transform
over $GF(2^m)$ is $n = 2^m - 1$. 

Let us consider cyclotomic cosets modulo
$n=2^m-1$ over $GF(2)$ 

\begin{eqnarray*}
\{c_0\} = \{0\}\\
 \{c_1,c_1 2,c_1 2^2, \ldots,c_1 2^{m_1-1}\},\\
\ldots,\\
 \{c_l,c_l 2,c_l 2^2, \ldots,c_l 2^{m_l-1}\},
\end{eqnarray*}
where $c_k \equiv c_k 2^{m_k} \bmod n$,
$l+1$ is the number of cyclotomic cosets modulo $n$ over $GF(2)$.

Let us introduce the set of indices modulo $n$

$$
Z = (Z_i) = (c_0,c_1,c_12,c_12^2, \ldots,c_12^{m_1-1}, \ldots,$$
$$c_l,c_l2,c_l2^2, \ldots,c_l2^{m_l-1}), \quad i \in [0,n-1].
$$

Then, we define a permutation matrix $\Pi=(\Pi_{i,j})$, \, $i,j \in [0,n-1]$,

$$ 
\Pi_{i,j} = 
\left\{
\begin{matrix}
1, & \hbox{ if } j = Z_i, \quad i \in [0,n-1] \cr
0, & \hbox{ otherwise}.                    \cr
\end{matrix}
\right.
$$

Let us denote a basis
$\beta_k=\left(\beta_{k,0}, \ldots ,\beta_{k,m_k-1}\right)$ of the subfield $GF(2^{m_k}) \subset GF(2^m)$.

Then we can write the cyclotomic DFT \cite{Fedorenko03,Fedorenko04}

\begin{equation*}
F_j=\sum_{k=0}^l\sum_{s=0}^{m_k-1}a_{k,j,s}\left(\sum_{p=0}^{m_k-1}
\beta_{k,s}^{2^p}f_{c_k 2^p}\right),
\end{equation*}
where $a_{k,j,s}\in GF(2)$. 

This equation can be represented in matrix form as

\begin{equation}
F=AL (\Pi f),
\end{equation} 
where $A$ is a matrix with elements $a_{k,j,s}\in GF(2)$ 
and $L$ is a block diagonal matrix with elements $\beta_{k,s}^{2^p}$.
If one chooses the normal basis $\beta_k$, then all the blocks of 
the matrix $L$ are circulant matrices. 

The inverse DFT in the field $GF(2^m)$ is  
$$
f=W^{-1}F.
$$

It is easily shown that 
\begin{equation}
W^{-1} = E W,
\end{equation} 
where $E$ is a matrix

$$ 
E =
\begin{pmatrix}
1     & 0     & 0    & \cdot & 0    & \cdot & 0    & 0     \cr
0     & 0     & 0    & \cdot & 0    & \cdot & 0    & 1     \cr
0     & 0     & 0    & \cdot & 0    & \cdot & 1    & 0     \cr
\cdot & \cdot &\cdot & \cdot & \cdot& \cdot & \cdot& \cdot \cr
0     & 0     & 0    & \cdot & 1    & \cdot & 0    & 0     \cr
\cdot & \cdot &\cdot & \cdot & \cdot& \cdot & \cdot& \cdot \cr
0     & 0     & 1    & \cdot & 0    & \cdot & 0    & 0     \cr
0     & 1     & 0    & \cdot & 0    & \cdot & 0    & 0     \cr
\end{pmatrix}.
$$

\bigskip

\section{THE INVERSE CYCLOTOMIC DFT}

Since both matrices $A$ and $L$ are invertible, from (2) 
the following representation of the inverse DFT can be derived
\begin{equation*}
\Pi f = L^{-1} A^{-1} F.
\end{equation*}

\begin{Lemma}[\cite{Hong}]
\label{lemma1}
Suppose $\beta_k$ are the normal bases,
then it is possible to show that blocks of $L^{-1}$ consist
of elements of bases $\beta_{k}'$ which are dual to $\beta_{k}$,
that is, the blocks of $L^{-1}$ are also circulant matrices. 
\end{Lemma}

\begin{Theorem}
\label{theorem1}
The inverse cyclotomic DFT for $GF(2^m)$ is
$$
(\Pi E) F = L^{-1} A^{-1} f.
$$
\end{Theorem}

\begin{IEEEproof}
From (1) and (2) we have 

$$
W = AL\Pi
$$ 
and 
$$
W^{-1} = \Pi^{-1} L^{-1} A^{-1}.
$$

From the last formula and (3) we obtain 
$$
W^{-1} = E W = \Pi^{-1} L^{-1} A^{-1} 
$$
and
$$
W = E^{-1} \Pi^{-1} L^{-1} A^{-1}. 
$$

Hence,
$$
F = W f = (E^{-1} \Pi^{-1}) L^{-1} A^{-1} f.
$$
and
$$
(\Pi E) F = L^{-1} A^{-1} f.
$$
\end{IEEEproof}

\section{EXAMPLES}

\subsection{DFT of length $n=7$}

Let $(\gamma,\gamma^2,\gamma^4)$ be a normal basis of $GF(2^3)$, 
where $\gamma=\alpha^3$ and $\alpha$ is a root of the primitive polynomial $x^3+x+1$.  
Then the cyclotomic DFT can be represented as

$$
F =
\begin{pmatrix}
F_0\\
F_1\\
F_2\\
F_3\\
F_4\\
F_5\\
F_6\\
\end{pmatrix}=
\begin{pmatrix}
1&1&1&1&1&1&1\\
1&0&1&1&1&0&0\\
1&1&0&1&0&1&0\\
1&1&0&0&1&0&1\\
1&1&1&0&0&0&1\\
1&0&0&1&0&1&1\\
1&0&1&0&1&1&0
\end{pmatrix}
$$
$$
\times
\begin{pmatrix}
1&0       &0       &0         &0       &0       &0       \\
0&\gamma^1&\gamma^2&\gamma^4  &0       &0       &0       \\
0&\gamma^2&\gamma^4&\gamma^1  &0       &0       &0       \\
0&\gamma^4&\gamma^1&\gamma^2  &0       &0       &0       \\
0&0       &0       &0         &\gamma^1&\gamma^2&\gamma^4\\
0&0       &0       &0         &\gamma^2&\gamma^4&\gamma^1\\
0&0       &0       &0         &\gamma^4&\gamma^1&\gamma^2
\end{pmatrix}
$$
$$
\times
\begin{pmatrix}
 1 &0 &0 &0 &0 &0 &0 \\
 0 &1 &0 &0 &0 &0 &0 \\
 0 &0 &1 &0 &0 &0 &0 \\
 0 &0 &0 &0 &1 &0 &0 \\
 0 &0 &0 &1 &0 &0 &0 \\
 0 &0 &0 &0 &0 &0 &1 \\
 0 &0 &0 &0 &0 &1 &0 \\
\end{pmatrix}
\begin{pmatrix}
f_0\\
f_1\\
f_2\\
f_3\\
f_4\\
f_5\\
f_6\\
\end{pmatrix}
= AL \Pi f.
$$

Using Theorem~\ref{theorem1}, we obtain the inverse cyclotomic DFT 

$$
(\Pi E) F
$$
$$ =
\begin{pmatrix}
 1 &0 &0 &0 &0 &0 &0 \\
 0 &1 &0 &0 &0 &0 &0 \\
 0 &0 &1 &0 &0 &0 &0 \\
 0 &0 &0 &0 &1 &0 &0 \\
 0 &0 &0 &1 &0 &0 &0 \\
 0 &0 &0 &0 &0 &0 &1 \\
 0 &0 &0 &0 &0 &1 &0 \\
\end{pmatrix}
\begin{pmatrix}
 1 &0 &0 &0 &0 &0 &0 \\
 0 &0 &0 &0 &0 &0 &1 \\
 0 &0 &0 &0 &0 &1 &0 \\
 0 &0 &0 &0 &1 &0 &0 \\
 0 &0 &0 &1 &0 &0 &0 \\
 0 &0 &1 &0 &0 &0 &0 \\
 0 &1 &0 &0 &0 &0 &0 \\
\end{pmatrix}
$$
$$
\times
\begin{pmatrix}
F_0\\
F_1\\
F_2\\
F_3\\
F_4\\
F_5\\
F_6\\
\end{pmatrix}
=
\begin{pmatrix}
1&0       &0       &0         &0       &0       &0       \\
0&\gamma^1&\gamma^2&\gamma^4  &0       &0       &0       \\
0&\gamma^2&\gamma^4&\gamma^1  &0       &0       &0       \\
0&\gamma^4&\gamma^1&\gamma^2  &0       &0       &0       \\
0&0       &0       &0         &\gamma^1&\gamma^2&\gamma^4\\
0&0       &0       &0         &\gamma^2&\gamma^4&\gamma^1\\
0&0       &0       &0         &\gamma^4&\gamma^1&\gamma^2
\end{pmatrix}
$$
$$
\times
\begin{pmatrix}
1&1&1&1&1&1&1\\
1&0&0&1&1&1&0\\
1&1&0&1&0&0&1\\
1&0&1&0&0&1&1\\
1&1&0&0&1&0&1\\
1&1&1&0&0&1&0\\
1&0&1&1&1&0&0
\end{pmatrix}
\begin{pmatrix}
f_0\\
f_1\\
f_2\\
f_3\\
f_4\\
f_5\\
f_6\\
\end{pmatrix}
= L^{-1} A^{-1} f.
$$

Finally note that the last formula coincides with formula \hbox{\cite[(6)]{Fedorenko04}}.

$$
\begin{pmatrix}
F_0\\
F_6\\
F_5\\
F_3\\
F_4\\
F_1\\
F_2\\
\end{pmatrix}
= L^{-1} A^{-1}
\begin{pmatrix}
f_0\\
f_1\\
f_2\\
f_3\\
f_4\\
f_5\\
f_6
\end{pmatrix}.
$$

This algorithm requires 6 multiplications and 
24 additions and appears to be the best known 7-point DFT for $GF(2^3)$.

\subsection{DFT of length $n=15$}

Let $(\alpha^{ 3},\alpha^{ 6},\alpha^{12},\alpha^{ 9})$ and $(\alpha^{11},\alpha^{ 7},\alpha^{14},\alpha^{13})$
be the normal bases of $GF(2^3)$, 
where $\alpha$ is a root of the primitive polynomial $x^4+x+1$.  
Then the cyclotomic DFT can be represented as formula (4) or $F=AL (\Pi f)$.
The inverse cyclotomic DFT can be written as formula (5) or $(\Pi E) F = L^{-1} A^{-1} f$.

\section*{ACKNOWLEDGMENT}

The author would like to thank Alexey Maevskiy for pointing out a methodological mistake in the paper \cite{Fedorenko04},
and Peter Trifonov for helpful discussions.


\begin{figure*}[!t]
\normalsize
$$
\begin{pmatrix}
F_0\\
F_1\\
F_2\\
F_3\\
F_4\\
F_5\\
F_6\\
F_7\\
F_8\\
F_9\\
F_{10}\\
F_{11}\\
F_{12}\\
F_{13}\\
F_{14}\\
\end{pmatrix}=
\begin{pmatrix}
  1  1  1  1  1  1  1  1  1  1  1  1  1  1  1 \\
  1  1  0  0  1  1  0  0  0  1  1  1  0  1  0 \\
  1  1  1  0  0  0  1  0  0  0  1  1  1  0  1 \\
  1  1  0  0  0  0  0  0  1  0  1  0  0  1  1 \\
  1  0  1  1  0  0  0  1  0  1  0  1  1  1  0 \\
  1  0  1  0  1  1  1  1  1  0  1  0  1  0  1 \\
  1  0  1  0  0  1  0  0  0  0  0  1  0  1  1 \\
  1  1  1  1  0  0  1  0  0  0  1  1  0  1  0 \\
  1  0  0  1  1  0  0  0  1  1  1  0  1  0  1 \\
  1  0  0  0  1  0  0  1  0  1  0  0  0  1  1 \\
  1  1  0  1  0  1  1  1  1  1  0  1  0  1  0 \\
  1  1  1  0  1  1  0  0  0  1  1  0  0  0  1 \\
  1  0  0  1  0  0  1  0  0  0  0  0  1  1  1 \\
  1  1  0  1  1  0  0  0  1  1  0  0  1  1  0 \\
  1  0  1  1  1  0  0  1  0  0  0  1  1  0  1 \\
\end{pmatrix}
$$
$$ 
\times
\begin{pmatrix}
\begin{tabular}{lllllllllllllll} 
  1 &0             &0             &0             &0             &0             &0             &0             &0             &0             &0             &0             &0              &0             &0            \\
  0 &$\alpha^{ 3}$ &$\alpha^{ 6}$ &$\alpha^{12}$ &$\alpha^{ 9}$ &0             &0             &0             &0             &0             &0             &0             &0              &0             &0            \\
  0 &$\alpha^{ 6}$ &$\alpha^{12}$ &$\alpha^{ 9}$ &$\alpha^{ 3}$ &0             &0             &0             &0             &0             &0             &0             &0              &0             &0            \\
  0 &$\alpha^{12}$ &$\alpha^{ 9}$ &$\alpha^{ 3}$ &$\alpha^{ 6}$ &0             &0             &0             &0             &0             &0             &0             &0              &0             &0            \\
  0 &$\alpha^{ 9}$ &$\alpha^{ 3}$ &$\alpha^{ 6}$ &$\alpha^{12}$ &0             &0             &0             &0             &0             &0             &0             &0              &0             &0            \\
  0 &0             &0             &0             &0             &$\alpha^{ 3}$ &$\alpha^{ 6}$ &$\alpha^{12}$ &$\alpha^{ 9}$ &0             &0             &0             &0              &0             &0            \\
  0 &0             &0             &0             &0             &$\alpha^{ 6}$ &$\alpha^{12}$ &$\alpha^{ 9}$ &$\alpha^{ 3}$ &0             &0             &0             &0              &0             &0            \\
  0 &0             &0             &0             &0             &$\alpha^{12}$ &$\alpha^{ 9}$ &$\alpha^{ 3}$ &$\alpha^{ 6}$ &0             &0             &0             &0              &0             &0            \\
  0 &0             &0             &0             &0             &$\alpha^{ 9}$ &$\alpha^{ 3}$ &$\alpha^{ 6}$ &$\alpha^{12}$ &0             &0             &0             &0              &0             &0            \\
  0 &0             &0             &0             &0             &0             &0             &0             &0             &$\alpha^{ 3}$ &$\alpha^{ 6}$ &$\alpha^{12}$ &$\alpha^{ 9}$  &0             &0            \\
  0 &0             &0             &0             &0             &0             &0             &0             &0             &$\alpha^{ 6}$ &$\alpha^{12}$ &$\alpha^{ 9}$ &$\alpha^{ 3}$  &0             &0            \\
  0 &0             &0             &0             &0             &0             &0             &0             &0             &$\alpha^{12}$ &$\alpha^{ 9}$ &$\alpha^{ 3}$ &$\alpha^{ 6}$  &0             &0            \\
  0 &0             &0             &0             &0             &0             &0             &0             &0             &$\alpha^{ 9}$ &$\alpha^{ 3}$ &$\alpha^{ 6}$ &$\alpha^{12}$  &0             &0            \\
  0 &0             &0             &0             &0             &0             &0             &0             &0             &0             &0             &0             &0              &$\alpha^{ 5}$ &$\alpha^{10}$\\
  0 &0             &0             &0             &0             &0             &0             &0             &0             &0             &0             &0             &0              &$\alpha^{10}$ &$\alpha^{ 5}$\\
\end{tabular}
\end{pmatrix}
$$

\begin{equation}
\times
\begin{pmatrix}
 1 0 0 0 0 0 0 0 0 0 0 0 0 0 0 \\
 0 1 0 0 0 0 0 0 0 0 0 0 0 0 0 \\
 0 0 1 0 0 0 0 0 0 0 0 0 0 0 0 \\
 0 0 0 0 1 0 0 0 0 0 0 0 0 0 0 \\
 0 0 0 0 0 0 0 0 1 0 0 0 0 0 0 \\
 0 0 0 1 0 0 0 0 0 0 0 0 0 0 0 \\
 0 0 0 0 0 0 1 0 0 0 0 0 0 0 0 \\
 0 0 0 0 0 0 0 0 0 0 0 0 1 0 0 \\
 0 0 0 0 0 0 0 0 0 1 0 0 0 0 0 \\
 0 0 0 0 0 0 0 1 0 0 0 0 0 0 0 \\
 0 0 0 0 0 0 0 0 0 0 0 0 0 0 1 \\
 0 0 0 0 0 0 0 0 0 0 0 0 0 1 0 \\
 0 0 0 0 0 0 0 0 0 0 0 1 0 0 0 \\
 0 0 0 0 0 1 0 0 0 0 0 0 0 0 0 \\
 0 0 0 0 0 0 0 0 0 0 1 0 0 0 0 \\
\end{pmatrix}
\begin{pmatrix}
f_0\\
f_1\\
f_2\\
f_3\\
f_4\\
f_5\\
f_6\\
f_7\\
f_8\\
f_9\\
f_{10}\\
f_{11}\\
f_{12}\\
f_{13}\\
f_{14}\\
\end{pmatrix}
\end{equation}
\vspace*{4pt}
\end{figure*}

\begin{figure*}[!t]
\normalsize
$$
\begin{pmatrix}
 1 0 0 0 0 0 0 0 0 0 0 0 0 0 0 \\
 0 1 0 0 0 0 0 0 0 0 0 0 0 0 0 \\
 0 0 1 0 0 0 0 0 0 0 0 0 0 0 0 \\
 0 0 0 0 1 0 0 0 0 0 0 0 0 0 0 \\
 0 0 0 0 0 0 0 0 1 0 0 0 0 0 0 \\
 0 0 0 1 0 0 0 0 0 0 0 0 0 0 0 \\
 0 0 0 0 0 0 1 0 0 0 0 0 0 0 0 \\
 0 0 0 0 0 0 0 0 0 0 0 0 1 0 0 \\
 0 0 0 0 0 0 0 0 0 1 0 0 0 0 0 \\
 0 0 0 0 0 0 0 1 0 0 0 0 0 0 0 \\
 0 0 0 0 0 0 0 0 0 0 0 0 0 0 1 \\
 0 0 0 0 0 0 0 0 0 0 0 0 0 1 0 \\
 0 0 0 0 0 0 0 0 0 0 0 1 0 0 0 \\
 0 0 0 0 0 1 0 0 0 0 0 0 0 0 0 \\
 0 0 0 0 0 0 0 0 0 0 1 0 0 0 0 \\
\end{pmatrix}
\begin{pmatrix}
 1 0 0 0 0 0 0 0 0 0 0 0 0 0 0 \\
 0 0 0 0 0 0 0 0 0 0 0 0 0 0 1 \\
 0 0 0 0 0 0 0 0 0 0 0 0 0 1 0 \\
 0 0 0 0 0 0 0 0 0 0 0 0 1 0 0 \\
 0 0 0 0 0 0 0 0 0 0 0 1 0 0 0 \\
 0 0 0 0 0 0 0 0 0 0 1 0 0 0 0 \\
 0 0 0 0 0 0 0 0 0 1 0 0 0 0 0 \\
 0 0 0 0 0 0 0 0 1 0 0 0 0 0 0 \\
 0 0 0 0 0 0 0 1 0 0 0 0 0 0 0 \\
 0 0 0 0 0 0 1 0 0 0 0 0 0 0 0 \\
 0 0 0 0 0 1 0 0 0 0 0 0 0 0 0 \\
 0 0 0 0 1 0 0 0 0 0 0 0 0 0 0 \\
 0 0 0 1 0 0 0 0 0 0 0 0 0 0 0 \\
 0 0 1 0 0 0 0 0 0 0 0 0 0 0 0 \\
 0 1 0 0 0 0 0 0 0 0 0 0 0 0 0 \\
\end{pmatrix}
\begin{pmatrix}
F_0\\
F_1\\
F_2\\
F_3\\
F_4\\
F_5\\
F_6\\
F_7\\
F_8\\
F_9\\
F_{10}\\
F_{11}\\
F_{12}\\
F_{13}\\
F_{14}\\
\end{pmatrix}
$$
$$
=
\begin{pmatrix}
\begin{tabular}{lllllllllllllll} 
  1 &0             &0             &0             &0             &0             &0             &0             &0             &0             &0             &0             &0              &0             &0            \\
  0 &$\alpha^{11}$ &$\alpha^{ 7}$ &$\alpha^{14}$ &$\alpha^{13}$ &0             &0             &0             &0             &0             &0             &0             &0              &0             &0            \\
  0 &$\alpha^{ 7}$ &$\alpha^{14}$ &$\alpha^{13}$ &$\alpha^{11}$ &0             &0             &0             &0             &0             &0             &0             &0              &0             &0            \\
  0 &$\alpha^{14}$ &$\alpha^{13}$ &$\alpha^{11}$ &$\alpha^{ 7}$ &0             &0             &0             &0             &0             &0             &0             &0              &0             &0            \\
  0 &$\alpha^{13}$ &$\alpha^{11}$ &$\alpha^{ 7}$ &$\alpha^{14}$ &0             &0             &0             &0             &0             &0             &0             &0              &0             &0            \\
  0 &0             &0             &0             &0             &$\alpha^{11}$ &$\alpha^{ 7}$ &$\alpha^{14}$ &$\alpha^{13}$ &0             &0             &0             &0              &0             &0            \\
  0 &0             &0             &0             &0             &$\alpha^{ 7}$ &$\alpha^{14}$ &$\alpha^{13}$ &$\alpha^{11}$ &0             &0             &0             &0              &0             &0            \\
  0 &0             &0             &0             &0             &$\alpha^{14}$ &$\alpha^{13}$ &$\alpha^{11}$ &$\alpha^{ 7}$ &0             &0             &0             &0              &0             &0            \\
  0 &0             &0             &0             &0             &$\alpha^{13}$ &$\alpha^{11}$ &$\alpha^{ 7}$ &$\alpha^{14}$ &0             &0             &0             &0              &0             &0            \\
  0 &0             &0             &0             &0             &0             &0             &0             &0             &$\alpha^{11}$ &$\alpha^{ 7}$ &$\alpha^{14}$ &$\alpha^{13}$  &0             &0            \\
  0 &0             &0             &0             &0             &0             &0             &0             &0             &$\alpha^{ 7}$ &$\alpha^{14}$ &$\alpha^{13}$ &$\alpha^{11}$  &0             &0            \\
  0 &0             &0             &0             &0             &0             &0             &0             &0             &$\alpha^{14}$ &$\alpha^{13}$ &$\alpha^{11}$ &$\alpha^{ 7}$  &0             &0            \\
  0 &0             &0             &0             &0             &0             &0             &0             &0             &$\alpha^{13}$ &$\alpha^{11}$ &$\alpha^{ 7}$ &$\alpha^{14}$  &0             &0            \\
  0 &0             &0             &0             &0             &0             &0             &0             &0             &0             &0             &0             &0              &$\alpha^{ 5}$ &$\alpha^{10}$\\
  0 &0             &0             &0             &0             &0             &0             &0             &0             &0             &0             &0             &0              &$\alpha^{10}$ &$\alpha^{ 5}$\\
\end{tabular}
\end{pmatrix}
$$

\begin{equation}
\times
\begin{pmatrix}
 1 1 1 1 1 1 1 1 1 1 1 1 1 1 1 \\
 1 0 0 0 1 1 1 1 0 1 0 1 1 0 0 \\
 1 0 0 1 0 0 0 1 1 1 1 0 1 0 1 \\
 1 1 0 1 0 1 1 0 0 1 0 0 0 1 1 \\
 1 0 1 1 0 0 1 0 0 0 1 1 1 1 0 \\
 1 0 1 1 1 1 0 1 1 1 1 0 1 1 1 \\
 1 1 0 1 1 1 1 0 1 1 1 1 0 1 1 \\
 1 1 1 1 0 1 1 1 1 0 1 1 1 1 0 \\
 1 1 1 0 1 1 1 1 0 1 1 1 1 0 1 \\
 1 1 0 1 0 1 1 1 1 0 0 0 1 0 0 \\
 1 1 1 0 0 0 1 0 0 1 1 0 1 0 1 \\
 1 0 1 1 1 1 0 0 0 1 0 0 1 1 0 \\
 1 0 0 1 1 0 1 0 1 1 1 1 0 0 0 \\
 1 0 1 1 0 1 1 0 1 1 0 1 1 0 1 \\
 1 1 0 1 1 0 1 1 0 1 1 0 1 1 0 \\
\end{pmatrix}
\begin{pmatrix}
f_0\\
f_1\\
f_2\\
f_3\\
f_4\\
f_5\\
f_6\\
f_7\\
f_8\\
f_9\\
f_{10}\\
f_{11}\\
f_{12}\\
f_{13}\\
f_{14}\\
\end{pmatrix}
\end{equation}
\vspace*{4pt}
\end{figure*}


\begin{thebibliography}{9}

\bibitem{Fedorenko03}
P. V. Trifonov and S. V. Fedorenko.
A method for fast computation of the Fourier transform over a finite field.
{\em Problems of Information Transmission}, 
vol. 39, no. 3, pp. 231--238, 2003.
Translation of {\em Problemy Peredachi Informatsii}. 

\bibitem{Fedorenko04}
E. Costa, S. V. Fedorenko, and P. V. Trifonov.
On computing the syndrome polynomial in Reed--Solomon decoder. 
{\em European Transactions on Telecommunications},
vol. 15, no. 3, pp. 337--342, 2004.

\bibitem{Hong}
J. Hong and M. Vetterli.
Computing $m$ DFT's over $GF(q)$ with one DFT over $GF(q^m)$.
{\em IEEE Transactions on Information Theory}, 
vol. 39, no.~1, pp. 271--274, January 1993.

\end{thebibliography}
\end{document}